\documentclass[10pt,conference]{IEEEtran}
\IEEEoverridecommandlockouts
\usepackage{cite}
\usepackage{amsmath,amssymb,amsfonts}
\usepackage{algorithmic}
\usepackage{graphicx}
\usepackage{textcomp}
\usepackage{xcolor}
\usepackage{epstopdf}
\usepackage{caption}
\usepackage{subcaption}
\usepackage[hidelinks]{hyperref}
\def\BibTeX{{\rm B\kern-.05em{\sc i\kern-.025em b}\kern-.08em
    T\kern-.1667em\lower.7ex\hbox{E}\kern-.125emX}}

\usepackage[nogroupskip,nonumberlist,nopostdot]{glossaries}
\usepackage{glossary-mcols}
\newacronym{hpc}{HPC}{High Performance Computing}
\newacronym{fea}{FEA}{Finite Element Analysis}
\newacronym{fdm}{FDM}{Finite Differences Method}
\newacronym{fem}{FEM}{Finite Element Method}
\newacronym{fvm}{FVM}{Finite Volume Method}
\newacronym{pde}{PDE}{Partial Differential Equation}
\newacronym{ai}{AI}{Artificial Intelligence}
\newacronym{ksp}{KSP}{Krylov Subspace Methods}
\newacronym{dof}{DOF}{Degrees of Freedom}
\newacronym{ebe}{EBE}{Element-by-element}
\newacronym{mp}{MP}{Mixed-precision}
\newacronym{gpu}{GPU}{Graphics Processing Unit}
\newacronym{mpi}{MPI}{Message-passing Interface}
\newacronym{cpu}{CPU}{Central Processing Unit}
\newacronym{simd}{SIMD}{Single Instruction Multiple Data}
\newacronym{cg}{CG}{Conjugate Gradient}
\newacronym{pcg}{PCG}{Preconditioned Conjugate Gradient}
\newacronym{minres}{MINRES}{Minimal Residual Method}
\newacronym{gmres}{GMRES}{Generalised Minimal Residual Method}
\newacronym{bicg}{BiCG}{Biconjugate Gradient Method}
\newacronym{acg}{ACG}{Adaptive Conjugate Gradient}
\newacronym{spmv}{SpMV}{Sparse Matrix-Vector Multiplication}
\newacronym{spgemm}{SpGEMM}{Sparse Matrix-Matrix Multiplication}
\newacronym{amg}{AMG}{Algebraic Multi-grid}
\newacronym{gmg}{GMG}{Geometric Multi-grid}
\newacronym{dt}{DT}{Decision Tree}
\newacronym{cnn}{CNN}{Convolutional Neural Network}
\newacronym{ml}{ML}{Machine Learning}
\newacronym{dl}{DL}{Deep Learning}
\newacronym{sor}{SOR}{Successive Over-relaxation}
\newacronym{arm}{ARM}{Acorn RISC Machine}

\newacronym{coo}{COO}{Coordinate}
\newacronym{csr}{CSR}{Compressed Sparse Row}
\newacronym{dia}{DIA}{Diagonal}
\newacronym{ell}{ELL}{ELLPACK}
\newacronym{hyb}{HYB}{Hybrid}
\newacronym{hdc}{HDC}{Hybrid DIA/CSR}
\newacronym{hec}{HEC}{Hybrid ELL/CSR}

\newacronym{vtable}{vtable}{virtual function table}
\newacronym{hpcg-bench}{HPCG}{High Performance Conjugate Gradients}
\newacronym{hpl-bench}{HPL}{High Performance LINPACK}
\newacronym{hbm}{HBM}{High Bandwidth Memory}

\usepackage{lipsum}

\begin{document}

\title{Exploiting dynamic sparse matrices for performance portable linear algebra operations}

\author{\IEEEauthorblockN{Christodoulos Stylianou}
\IEEEauthorblockA{\textit{EPCC, The University of Edinburgh}\\
Edinburgh, UK \\
c.stylianou@ed.ac.uk}
\and
\IEEEauthorblockN{Mich\`{e}le Weiland}
\IEEEauthorblockA{\textit{EPCC, The University of Edinburgh}\\
Edinburgh, UK \\
m.weiland@epcc.ed.ac.uk}
}

\maketitle

\begin{abstract}
Sparse matrices and linear algebra are at the heart of scientific simulations. More than 70 sparse matrix storage formats have been developed over the years, targeting a wide range of hardware architectures and matrix types. Each format is developed to exploit the particular strengths of an architecture, or the specific sparsity patterns of matrices, and the choice of the right format can be crucial in order to achieve optimal performance. The adoption of dynamic sparse matrices that can change the underlying data-structure to match the computation at runtime without introducing prohibitive overheads has the potential of optimizing performance through dynamic format selection.

In this paper, we introduce \emph{Morpheus}, a library that provides an efficient abstraction for dynamic sparse matrices. The adoption of dynamic matrices aims to improve the productivity of developers and end-users who do not need to know and understand the implementation specifics of the different formats available, but still want to take advantage of the optimization opportunity to improve the performance of their applications. We demonstrate that by porting HPCG to use \emph{Morpheus}, and without further code changes, 1) HPCG can now target heterogeneous environments and 2) the performance of the \gls{spmv} kernel is improved up to $2.5\times$ and $7\times$ on CPUs and GPUs respectively, through runtime selection of the best format on each MPI process.
\end{abstract}

\begin{IEEEkeywords}
  sparse matrix storage formats, generic programming, dynamic matrices, performance portability, productivity
\end{IEEEkeywords}
\section{Introduction}
Sparse matrices (i.e. matrices that mostly consist of zeros) are an essential concept in computational science and engineering.\cite{Colella_2006} The computational and memory savings that can be gained from exploiting the sparsity structure of a matrix have driven the development of software specific to sparse linear algebra. As the majority of elements in a sparse matrix are zeros, a special focus was given to defining formats (i.e. data structures) to enable the efficient storage of, and access to, all non-zero elements without the need to explicitly store the zeros as well. Sparse matrix storage formats reduce the memory footprint of the matrix, eliminate redundant computations and allow for larger problems to be processed.

According to Filippone et al.\cite{spmv_gpu_review}, more than $70$ such sparse matrix storage formats
have been developed over the years to address not only the various types of matrices that result from different discretization methods, but also the evolution of hardware towards multi- and many-core processors and accelerators. Literature shows that there is no single format that can perform optimally across all different kinds of matrices and types of hardware~\cite{spmv_gpu_review,cnn_spmv,csr5,axt,sell_c_sigma}. Compared to their dense counterparts, operations on sparse matrices are generally known to be memory-bandwidth bound because of the need to retrieve matrix values via the index information, often resulting in indirect memory accesses and poor cache reuse. 

In many numerical applications, computing the solution to linear systems (through performing sparse matrix-vector multiplications) dominates the runtime. The (iterative) solvers can be optimized through the use of storage formats that exploit hardware capabilities given the sparsity pattern (i.e. the distribution of non-zero entries) of the matrix and the operation to be performed. To facilitate the selection of the best format for a given matrix and target architecture, auto-tuners have been developed~\cite{cnn_spmv, smat, clSpMV}. These auto-tuners demonstrate the impact of selecting a suitable format at runtime to optimise the performance of iterative solvers on single node multi- or many-core systems and they provide mechanisms to automate the process.

In this work, we develop an abstraction for sparse matrices that enables switching to different storage formats at runtime. This abstraction  provides optimization opportunities, by adapting the data-structure given the operation, target architecture and sparsity pattern of the matrix, but without introducing noticeable overheads. We introduce \emph{Morpheus}\cite{morpheus}, a library that supports the runtime-switching of matrix storage formats by implementing sparse matrices using a single dynamic ``abstract'' format and providing a transparent mechanism to switch between different implementations. A detailed description of \emph{Morpheus} is given in Section~\ref{sec:morpheus}. We use the \gls{hpcg-bench} benchmark\cite{hpcg} as a test case to represent a common numerical workload, and we extend it to support dynamic sparse matrix storage format switching and multiple hardware backends that makes it possible to target CPU or GPU architectures without any further code modifications.

In summary, our contributions are:
\begin{itemize}
    \item We show that using the abstract sparse matrix format representation provided by \emph{Morpheus}\cite{morpheus}, our library of sparse matrix storage formats, incurs no significant runtime overheads. Indeed, without any further code changes and by transparently switching to a different format, performance of a solver can often improve (see Section~\ref{sec:overheads}).
    \item We describe a general process for incremental porting of an application to use \emph{Morpheus} (see Section~\ref{sec:porting-generic}) and, as validation, we provide an example of the process for HPCG (see Sections~\ref{sec:porting-hpcg}).
    \item We demonstrate that for an optimal storage format, a specific problem in a shared memory setting can introduce irregularities in the sparsity pattern and perform significantly worse for the same format in a distributed memory setting. We show that this issue can be solved by splitting the local system matrix into a local (regular) and remote (irregular) part, each with possibly different formats, and investigate the performance benefits from doing so.
    \item We show that by selecting the optimal format for local and remote matrices per process the application can achieve and retain the best possible runtime performance. After enabling HPCG to use \emph{Morpheus}, we are able to perform the \gls{spmv} routine up to $2.5\times$ faster on a multi-core CPU system and up to $7\times$ faster on CPU+GPU nodes, using the same source code and without any further modifications.
\end{itemize}


\section{Background and Motivation}

\subsection{Sparse Matrix Storage Formats}\label{sec:sparse_matrix_storage_formats}

Dense matrices store all the coefficients of a matrix explicitly in memory, usually by using a two-dimensional linear mapping. The coefficient $A(i,j)$ of a dense $M \times N$ matrix is stored at position $(i \times N) + j$ of a linear array (assuming row-major storage ordering). Sparse matrix formats on the other hand exploit the property that the majority of sparse matrix coefficients are zeros by not explicitly storing those values. As a result, the direct link between the index pair $(i,j)$ and the position of the coefficient in memory is lost. A sparse matrix storage format aims to rebuild this link using auxiliary index information. The cost of calculating this indirection has an impact on the performance of sparse matrix computations. 

A large number of sparse matrix storage formats have been developed over the years, each with different storage requirements, computational characteristics, and methods of modifying the entries of the matrix.\cite{segmented_scan} As a result, determining which format might perform best given an operation is not a trivial task as multiple factors determine the overall performance. No single storage format is able to exploit the matrix structure and perform optimally across multiple operations, or indeed across multiple target architectures.\cite{spmv_gpu_review}

The most basic and well-known formats are \gls{coo} and \gls{csr}. Both are considered \emph{general purpose} formats, suitable for a broad range of matrices of arbitrary sparsity patterns and target architectures. \gls{coo} uses three arrays, where each non-zero element is explicitly stored together with its column and row indices, with no guarantees imposed in the ordering of the elements. \gls{csr} also explicitly stores the column indices and non-zero values, but it uses an array of pointers to mark the boundaries of each row, reducing the memory footprint of the format by essentially compressing the row indices. As the row pointers are used to represent the position of the first non-zero element in each row, and the last entry shows the total number of non-zeros in the matrix, \gls{csr} naturally also imposes an ordering across rows, but not within each row. 

\emph{Specific purpose} formats aim to address the characteristics of specific classes of matrices and are usually designed to perform optimally with a target architecture in mind. For example, the \gls{dia} format was originally designed for vector processors and is suitable for regular sparsity patterns. \gls{dia} uses a two-dimensional array, where each column holds the coefficients of a diagonal of the matrix, and an integer offset array keeping track of where each diagonal starts. Therefore, \gls{dia} format is suitable for matrices with structures that dominate along the diagonals, such as banded matrices resulting from discretization methods like \gls{fdm}, and vector-like architectures, such as GPUs.

Note that the formats above, or a combination of them, constitute a basis from where many other formats are derived. However, it is clear that the underlying data structure across formats can vary significantly. As a result, accessing and manipulating entries in each format can result in different memory access patterns, costs and interfaces amongst formats.

\subsection{HPCG}\label{sec::hpcg}

\gls{hpcg-bench}\cite{hpcg} is a benchmark that measures the performance of HPC systems by solving the Poisson differential equation on a regular 3D grid, discretized with a 27-point stencil. It uses the \gls{pcg} algorithm with a symmetric Gauss-Seidel\cite{Saad_2003} as a preconditioner, and includes the following computations: 
    sparse matrix-vector multiplications (\glspl{spmv});
    vector updates;
    global dot products;
    a local symmetric Gauss-Seidel smoother (including a sparse triangular solve);
    and multi-grid (MG) preconditioned solvers.
The benchmark compares and validates the performance of a user-tunable optimized version of the \gls{pcg} algorithm against a reference in the following phases:
\begin{enumerate}
    \item \textit{Problem setup}: Constructs the synthetic problem by creating the geometry and linear system.
    \item \textit{Reference timing}: Measures the time taken to run the \gls{spmv} and MG reference implementations and the time to solution for the reference \gls{cg} solver.
    \item \textit{Problem Optimization setup}: Configures the user defined data structures to be used in the optimized problem.
    \item \textit{Validation and Verification}: Checks that the optimized problem has returned the expected results.
    \item \textit{Optimized problem timing}: Measures the time to solution for the optimized \gls{cg} solver.
\end{enumerate}
We choose \gls{hpcg-bench} here because, not only it is a widely accepted and well-understood benchmark, but also because it is designed to use optimized implementations (in our case provided by \emph{Morpheus}) of the linear algebra computations and compare them against a reference baseline. Note that the reference implementation of \gls{hpcg-bench} can be built with MPI for distributed parallelism and/or OpenMP for shared memory parallelism.

The algorithm chosen for solving the system of equations is not only limited by the floating point performance but also heavily relies on the performance of the memory system and to some extent the underlying network. Note that similar computations and data access patterns are found in many scientific calculations. Therefore, \gls{hpcg-bench} provides a good performance assessment that is representative of the performance of many scientific applications.
The performance bottleneck in \gls{hpcg-bench}
boils down to the sparse operations that are carried out at every step of the iterative solver, i.e the \gls{spmv} and the Gauss-Seidel smoother.

\subsection{Motivation}

Scientific codes that use sparse matrices are often built around a single general-purpose storage format. However different applications and problems, including those from new application domains, will exhibit different matrix sparsity patterns. In addition, the continuous evolution of hardware introduces architectures with new characteristics. For software to remain performant throughout its lifetime, algorithms must be adjusted to exploit these new architectures and represent sparsity patterns efficiently. Previous work (such as~\cite{cnn_spmv, smat, clSpMV}) has demonstrated the importance of choosing the correct storage format in order to achieve optimal performance. However, only the impact on performance at the shared-memory was considered, without providing performance results at a distributed level. In a distributed environment, the global matrix is partitioned into smaller local matrices that are each assigned to a different process. As a result, the sparsity pattern of each local matrix can differ significantly. In this case, choosing a single storage format based on metrics that arise from the global matrix could result in noticeable load imbalances. This observation motivates a \emph{per-process} selection of the optimum format, with the possibility of using different formats across processes. \emph{Morpheus} supports both single and multi-format runtime configurations and as such can be used to compare both strategies and assess their performance and viability.

\section{Morpheus - a library for dynamic sparse matrices}
\label{sec:morpheus}
Adopting an abstraction that can change the underlying sparse matrix storage format efficiently at runtime has the potential of offering performance benefits without invasive code modifications. To achieve this, we have developed \emph{Morpheus}\footnote{Available at: https://github.com/morpheus-org/morpheus.git}, a C++ library that offers transparent, efficient and low overhead run-time switching between different sparse matrix storage formats across different target architectures. By abstracting the different formats under a single \emph{dynamic} format and encapsulating the internal implementation details, end-users are left with a simple and intuitive interface that abstracts away the complexities. At the same time, performance can be optimised by transparently switching to the ``best'' format for each operation, sparsity pattern and target hardware, without any code modifications.

The \emph{Morpheus} library follows a functional design and separates the data structures (containers) from the functions (algorithms), with the algorithms acting on the containers. To enable support for various hardware platforms and memory hierarchies, we are adopting the three core abstractions concepts offered by Kokkos\cite{kokkos3}: 
\begin{itemize}
    \item \emph{Execution Space}: Specifies \textit{where} the code will be executed. Examples include the CPU and GPU cores.
    \item \emph{Memory Space}: Specifies \textit{where} the data will reside in memory, enabling exploitation of the different characteristics of various types of memory, such as non-volatile memory or \gls{hbm}.
    \item \emph{Memory Layout}: Specifies \textit{how} the data will reside in memory, enabling efficient data access patterns in algorithm and architecture dependent optimisations.
\end{itemize}
The adoption of these abstractions facilitates the formulation of generic algorithms and containers that can be efficiently mapped to the different types of architectures and memory hierarchies that are currently available, but also enable the addition of future developments.

\begin{figure}[t]
    \centering
    \includegraphics[width=\columnwidth]{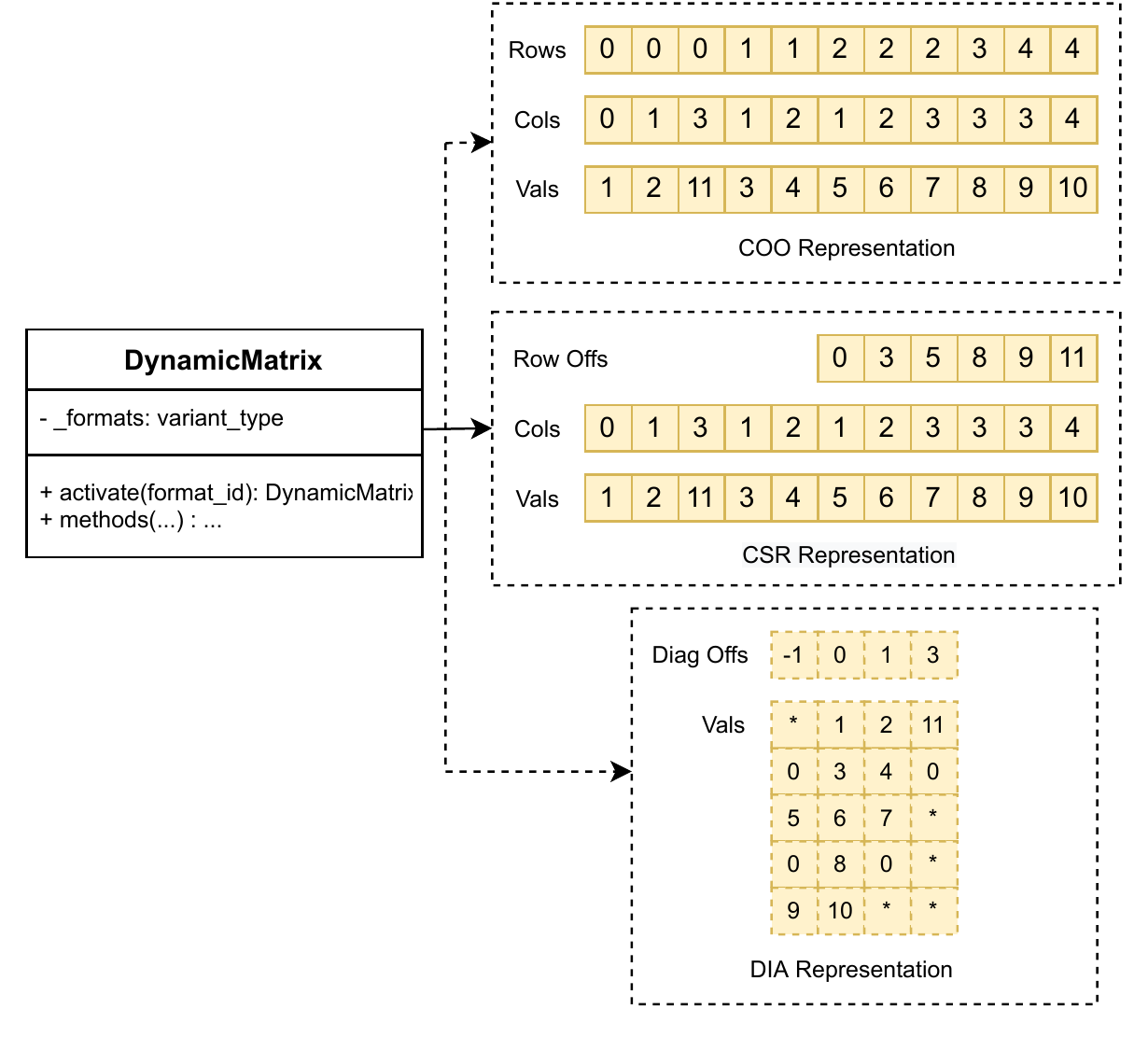}
    \setlength{\belowcaptionskip}{-8pt} 
    \caption{\texttt{DynamicMatrix} Container with CSR representation as its current active type (solid arrow). Other possible active types are represented via dashed arrows and can be switched to during run-time using the \emph{activate()} method.}
    \label{fig:dynamic_matrix}
\end{figure}

\subsection{Defining the data structures} 
Currently, \emph{Morpheus} supports three containers for sparse matrix storage formats (\emph{CooMatrix}, \emph{CsrMatrix} and \emph{DiaMatrix}), hereafter also called concrete formats, and two containers representing dense formats (\emph{DenseMatrix} and \emph{DenseVector}). Each container can be seen as a different type, uniformly parameterized by the type of the values and indices it holds, as well as where and how it resides in memory. Note that all the aforementioned containers are resolved at compile-time, and they can potentially each have a different interface.

To enable dynamic switching between formats we have introduced the concept of a \emph{DynamicMatrix}, a container acting as a composition of all the supported sparse containers. The \emph{DynamicMatrix} follows the same semantics and is parameterized in the same way as the concrete formats. A high-level overview of the \emph{DynamicMatrix} container is shown in Figure~\ref{fig:dynamic_matrix}. Through its own interface and the use of the \emph{State} pattern \cite{design_patterns}, which allows an object to alter its behaviour when its internal state changes, the \emph{DynamicMatrix} provides a unified way of interacting with all the supported storage formats. In other words, at any given moment it can hold any of the available storage formats participating in the composition and can switch to a different one by changing its active state. We allow the active state to be changed through the \verb|activate()| member function by selecting one of the available \emph{enum} values associated with each of the available storage formats or their equivalent index.

\subsection{Data management}
Each container is responsible for acquiring and releasing its own resources when it goes out of scope in order to avoid memory leaks. Communication and data exchange between containers is achieved using three different types of \emph{copy} concepts, each with a different set of requirements to be met and each with a different associated cost.

\textbf{Shallow Copy} implies that no actual copy or data transfer is performed, but that instead containers share resources, with deallocation  taking place once all containers sharing those same resource go out of scope. \emph{Shallow} copy has the lowest cost of the three copy concepts we support, however to succeed it requires that the two containers are of the same type (i.e the same format, type of values and indices, memory space and layout) otherwise a compile-time error is triggered.

\textbf{Deep Copy} performs a bit-wise copy of the data from a source to a destination container, and no shared state is maintained between the containers. The cost of such an operation is much higher compared to a shallow copy and increases as the size of the data increases, therefore such an operation has to be as efficient as possible. As a consequence, for a deep copy to be issued the source and destination containers must be compatible. In other words, the two containers must have the same format, type of values and indices with identical memory layout and alignment although the memory space can differ ensuring that the deep copy is transformed into a \emph{memcopy} operation. At the same time, this also holds for data transfers across different spaces. We extend the mirroring interface of \emph{Kokkos} to support sparse matrices, making sure that we can create and allocate compatible containers.
    
\textbf{Convert} performs an element-wise copy between two containers. This operation requires only that the two containers are in the same memory space, without imposing any further restrictions on the format, the type of values and indices or the layout and alignment. Therefore, this enables conversions from one sparse format to another, but also between containers of the same format that are not compatible, although at a higher cost compared to a deep copy operation. To avoid the need to provide one conversion implementation for each pair of supported storage formats, which could rapidly become intractable, we use a proxy format policy where \gls{coo} format acts as an intermediate format (i.e. all conversions involve going from the existing format \emph{to} \gls{coo} and \emph{from} \gls{coo} to a new format). Note that this is simply a pragmatic choice in order to control the complexity of the \emph{Morpheus} code base; creating specific optimised conversion operations (say from \gls{dia} directly to \gls{csr}) is of course possible.

It is also worth pointing out that in the case of \emph{shallow} and \emph{deep} copy operations, the same semantics apply to both the concrete and dynamic containers. The only difference is that when at least one of the containers participating in the operation is a \emph{DynamicMatrix}, its active state must also match the other container's state.

\subsection{Supporting heterogeneous platforms}\label{sec::host_device_model}

On heterogeneous systems, it might be necessary to perform data transfers across memory spaces. We assume that we have compute units responsible for general housekeeping (host) and compute units for performing the core computation (device). For example, on a CPU+GPU system, the CPU acts as the host and is responsible for preparing and transfering the data to the device (in this case the GPU), which in turn will perform the computation.

We adopt a host-device model to support heterogeneous platforms. Containers in \emph{Morpheus} are by default assumed to live on the device space and offer a \emph{HostMirror} type that represents an equivalent and compatible mirror container residing in the \emph{HostSpace}. Through the mirroring interface we can allocate host containers to match the size of their device counterpart and the data transfers from and to the two types of containers are managed performing deep copies. However, in the case where the device container already resides on \emph{HostSpace} (for example if only the CPU backend is enabled) both the device and host containers are the same hence performing deep copies between the two would be a redundant and expensive operation. To avoid such overheads, \emph{Morpheus} is intelligent enough to transform the deep copy operations in shallow copies, without user having to do any code modifications.

\subsection{Supporting different algorithms and programming models}
The algorithms supported in \emph{Morpheus} are exposed to the user through a generic high-level interface. Each algorithm has a unified interface such that it can be used in the same way across different containers. To specify where the algorithm will be executed, an \emph{ExecSpace} parameter must be provided, which must be a valid execution space provided by \emph{Kokkos}. Using compile-time introspection, \emph{Morpheus} selects and dispatches the appropriate algorithm implementation for the containers and execution space provided by the user.

In the case where at least one of the containers is a \emph{DynamicMatrix}, the same high-level interface is still used as \emph{DynamicMatrix} follows the same semantics as the concrete formats. However, changing the active state of the \emph{DynamicMatrix} at run-time would result in a different algorithm implementation being dispatched under the hood. We achieve this by using the \emph{Visitor} pattern\cite{design_patterns}, where a ``visitor''operation is responsible for inspecting the current state of the variant and for dispatching the correct routine. Note that because \emph{DynamicMatrix} knows all the active states it can switch to at compile-time, all the versions of the algorithm are generated by the compiler \emph{a priori} resulting in efficient run-time dispatches.

Currently, we support algorithms for serial, OpenMP and CUDA backend environments, with each algorithm kernel explicitly implemented for each backend. We also provide a \emph{GenericSpace} wrapper that will dispatch the generic version of the algorithm written in \emph{Kokkos}. For each matrix container we support basic operations such as \texttt{SpMV}, \texttt{diagonal update} and \texttt{extraction}. In addition, we provide algorithms such as \texttt{dot}, \texttt{WAXPBY}, \texttt{reduction} and \texttt{scan} for the \emph{DenseVector} container. The current algorithms we support are commonly found in scientific applications that use iterative solvers and represent a minimum set of functionalities needed for us to evaluate \emph{Morpheus}. For more complicated workloads, for example those using preconditioners, further algorithms will need to be implemented, although all will follow the same interface and design principles as discussed above.

\section{Integrating Morpheus into an existing application}\label{sec:porting}
The main goal of \emph{Morpheus} is to increase the end-user's productivity and the performance of their applications through efficient format switching and selection without the user having to dive into the specifics of each supported storage format. For that to be possible, \emph{Morpheus} needs to be able to be integrated in another application following a straightforward and incremental porting process, which is described below. For completeness, we also describe our approach to adding \emph{Morpheus} to \emph{HPCG} and provide details on the challenges and issues faced as well as how our library can help future-proof applications and benchmarks.

\subsection{Porting process}\label{sec:porting-generic}
Integrating \emph{Morpheus} into an existing application can be done incrementally in 3 steps, as described below.

\textbf{Step 1: Converting user-defined data structures.} In order to use the algorithms provided by \emph{Morpheus} we must first convert any user-defined data structures to the containers supported by \emph{Morpheus}. By default, a container is responsible for managing (i.e. allocating) its own resources, however it can also be specified to be ``unmanaged'', i.e the allocation has to be provided by the user during construction. Consequently, we can convert user-defined data structures to \emph{Morpheus} containers by passing an allocation that is used by the structure to an unmanaged container. As a result, both share the same resources and the container will be aliasing the original data structure. Any updates will be directly reflected on both.

For array-like data structures allocated by the user creating an unmanaged \emph{Morpheus} container is particularly useful. However, in the case of sparse matrices, the user-defined structure potentially has a completely different representation in memory from what \emph{Morpheus} supports. As a result, constructing a \emph{Morpheus} container will be only possible through an element-wise conversion provided by the user.

After the user-defined data structures have been converted into \emph{Morpheus} containers, the \emph{Morpheus} algorithms can now be used. However, at this stage only the CPU backends can be used as the GPU backend must first be enabled as it requires extra data management (as described in Step 2).

\textbf{Step 2: Enabling GPU support.} \emph{Morpheus} will not handle any data transfers between two different memory spaces, therefore it is up to the user to manage data transfers from the CPU and GPU. However, \emph{Morpheus} provides a mirroring interface and copy semantics as described in Section \ref{sec::host_device_model} that allows the user to target heterogeneous platforms and handle data transfers in a constructive way. 

During this step, the user has to decide which of the containers will be used in an algorithm and which will be used for general housekeeping. Consequently, any container that will be used in an algorithm will be assumed to reside in the device space and any access to its data outside of the algorithm will have to be made through its equivalent \emph{HostMirror} container. After the user handles the data transfers between device containers and \emph{HostMirror} containers,  the application code can now run both on CPU and GPU without any further modifications and using a single source code.

\textbf{Step 3: Enabling dynamic switching.} At this point, the application is able to run with \emph{Morpheus} both on CPU and GPU backends, however no dynamic format switching functionality is available yet. In order to enable run-time polymorphism to switch to different formats the user needs to convert the concrete sparse matrix container into a \emph{DynamicMatrix}. This is facilitated through the \emph{convert} routine, which will switch and convert to the desirable storage format representation. No further changes are required as algorithms for both concrete and dynamic matrices have the same high-level interface. The run time switching functionality therefore allows the application to not only be future proof in terms of the hardware it can target, but also in terms of the sparsity patterns it can efficiently process as more storage formats are added to \emph{Morpheus}.

\subsection{Example: HPCG with Morpheus}\label{sec:porting-hpcg}
The HPCG benchmark, described in detail in Section \ref{sec::hpcg}, was chosen as an exemplar application to test \emph{Morpheus} as it is a representative (albeit simplified) version of a real-life workload, with well understood performance characteristics. 
The goal is to create a \emph{Morpheus}-enabled version of HPCG\footnote{Available at: https://github.com/morpheus-org/morpheus-hpcg.git} to 1) confirm the soundness of the porting process, 2) identify any obstacles and pitfalls and 3) assess the performance impact of using \emph{Morpheus} as part of HPCG. Note that we concentrate on the main computational bottleneck of the application, the \gls{spmv} operation, and therefore for simplicity we disable the preconditioner step.

The reference implementation of HPCG benchmark runs on CPUs only and can be configured to use MPI and OpenMP. The benchmark has four main data structures: \emph{Vector}, \emph{SparseMatrix}, \emph{CGData} and \emph{MGData}. Here, we are interested in the first three (as preconditioning is disabled). The sections below describe the code changes that are required to port the reference HPCG implementation to support \emph{Morpheus}.

\subsubsection{The Vector and CGData data structures}
The first step is to transform all the vector-vector operations in HPCG to the equivalent algorithms from \emph{Morpheus}. To do so, we need to transform the \emph{Vector} data structure to also hold the \emph{DenseVector} container. Because the memory layout of both \emph{Vector} and \emph{DenseVector} is the same, we can construct an unmanaged \emph{DenseVector} using the existing memory allocations. Therefore, during the \texttt{Problem Optimization Setup} phase of HPCG, for every vector (i.e. $b$, $x$ and $xexact$) used in the \texttt{Optimized problem timing} phase, we initialize the unmanaged \emph{DenseVector} by passing it the allocation to the data. 

Since both \emph{Vector} and \emph{DenseVector} are now aliasing the same memory, they can be used interchangeably throughout the application without any need for further data management. We can now update HPCG's \texttt{ComputeDot} and \texttt{ComputeWAXPBY} functions to invoke the \emph{Morpheus} algorithms for \texttt{dot} and \texttt{WAXPBY}.

\subsubsection{The SparseMatrix data structure}
Changing the \emph{SparseMatrix} data structure to also hold one of the sparse matrix containers offered by \emph{Morpheus} requires an element-wise conversion between HPCG's format of choice and the format of choice from \emph{Morpheus}, which can be any of the formats that are supported as we are not limited to a single format, as is the case for other approaches like SMAT\cite{smat}. However, since HPCG uses a variation of CSR format it makes sense that the format of choice for \emph{Morpheus} container is \emph{CsrMatrix}.

During the \texttt{Problem Optimization Setup} phase we perform the conversion between the \emph{SparseMatrix} and the \emph{Morpheus} container for the system matrix $A$. Note that the two matrix representations live in different allocations and the user needs to ensure data consistency. However, since HPCG builds the system matrix in the \texttt{Problem Setup} phase and there is no change to it after that, both matrices will have the same data during the \texttt{Optimized problem timing} phase.

At this point, the \emph{ComputeSPMV} routine in HPCG can be modified to invoke the \emph{multiply} algorithm from \emph{Morpheus}. Note that although the computation in the iterative solver returns the same results in the \emph{Morpheus}-enabled HPCG, as it stands the benchmark will report invalid results. HPCG performs tests to ensure the solver works as expected and one of tests (\texttt{TestCG}) involves modifying the elements of the matrix diagonal. To address this issue, we have to update the diagonal of the \emph{Morpheus} container during the test by invoking the equivalent algorithm.

\subsubsection{Enabling GPU support}
During the \texttt{Problem Optimization Setup} phase, for every \emph{DenseVector} and \emph{CsrMatrix} container the equivalent \emph{HostMirror} container needs to be created through the \emph{Morpheus} mirroring interface, which will be initialized using the data from the HPCG data structures. The data will be loaded onto the device via a deep copy from the \emph{HostMirror} container.

Since the benchmark repeats multiple \emph{CG} runs, and within each run updates the vector data, these data transfers need to be setup manually by the user using the \emph{Morpheus} copy interface, i.e. any \texttt{CopyVector} call needs to be substituted with \texttt{Morpheus::copy}. In the same way, every time a vector is zeroed using \texttt{ZeroVector} operation, the \texttt{Morpheus::assign} routine must be used instead. This will ensure that irrespective of which backend is enabled, the data will be handled properly for the end results to be valid and all from the same source code.

If MPI is enabled, HPCG assumes that for the \texttt{ComputeSPMV} operation the source vector is distributed across processes, therefore every time prior to performing the operation the remote vector parts have to be exchanged. When the GPU backend is enabled each \emph{MPI} process is responsible for a GPU and the vector is distributed across the different GPUs. This means that the \texttt{ExchangeHalo} routine must be adapted to copy the data from one GPU to another via the CPU again using the \emph{Morpheus} copy interface.

Once this step has been completed, the benchmark is able to work with any backend that is supported now (or will be in the future) that follows the \emph{Host-Device Model} without any further source code modifications.

\subsubsection{Enabling Dynamic Switching}
Up until this point, the benchmark is using the \emph{CsrMatrix} container directly. If we want to use the dynamic capabilities of \emph{Morpheus}, the \emph{DynamicMatrix} container needs to be used instead. We change the container held by the \emph{SparseMatrix} to be the \emph{DynamicMatrix} and during the \texttt{Problem Optimization Setup} we assign the converted \emph{CsrMatrix} to the \emph{DynamicMatrix}. This will result in a \emph{DynamicMatrix} with its active state being the CSR format. To change the active state (i.e. switch to a different format) we can now apply an in-place conversion using the \emph{convert} interface from \emph{Morpheus} and selecting the format to use at run time, either through the command line or an input file. This \emph{Morpheus}-enabled version of HPCG will be able to take advantage of any new storage formats to optimise the performance without any code modifications.


\section{Results and Evaluation}
\subsection{Experiments}\label{sec:hypotheses}
Using the dynamic capabilities of \emph{Morpheus} as part of an application should not introduce significant overheads. Note that \emph{Morpheus} knows \emph{a priori} which format representations are supported and as a result, the compiler can generate a complete set of the possible algorithms that can be efficiently dispatched at runtime. To assess and quantify the overheads introduced to HPCG after adding support for \emph{Morpheus},  we compare the runtime of the original HPCG with respect to the \emph{Morpheus}-enabled HPCG with the active state of the \emph{DynamicMatrix} set to CSR. We perform this comparison for a set of compilers.

\begin{figure}[t]
    \centering
    \includegraphics[width=0.8\columnwidth]{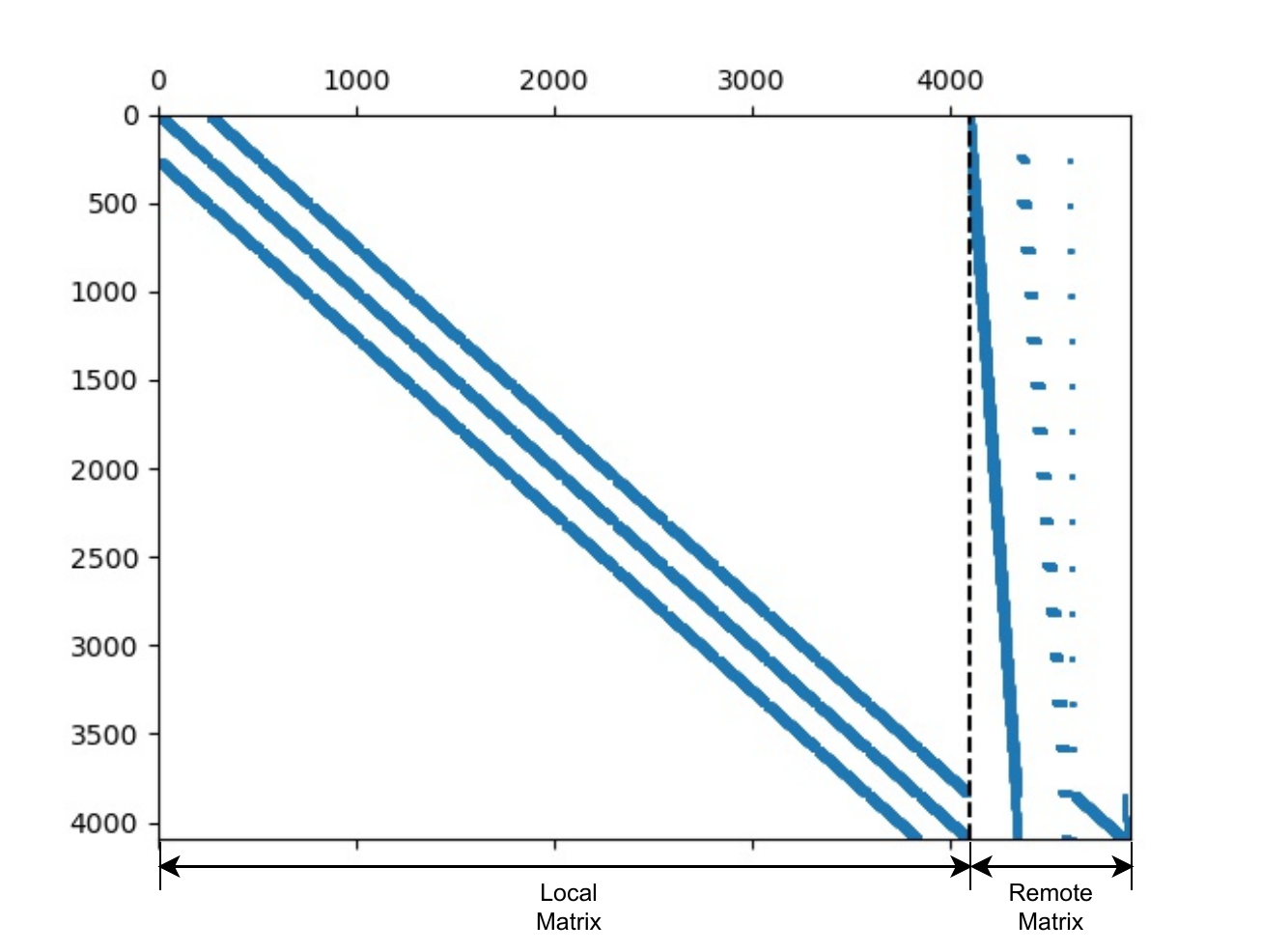}
    \setlength{\belowcaptionskip}{-8pt}     
    \caption{Image representation of the sparsity pattern of a 4096x4096 local system matrix at rank 0. The dashed line indicates which part of the system matrix belongs to the local matrix and which to the remote, demonstrating how the matrix will be divided into the two parts. Note that when MPI is disabled the remote matrix part disappears and the sparsity pattern of the local system matrix is regular and concentrated around the diagonals.}
    \label{fig:sparsity_matrix}
\end{figure}

Furthermore, since HPCG solves a Poisson differential equation on a regular 3D grid discretized with a 27-point stencil, the system matrix is expected to be regular with non-zeros concentrated around the diagonals. For that type of sparsity pattern the performance of the \gls{spmv} kernel is expected to be improved by switching to \gls{dia} format instead of \gls{csr}, as it will enable contiguous accesses to vector $x$. However, once MPI is enabled and in order to optimize the communication of the remote vector elements, on each process the local system matrix is conceptually divided into a local and a remote part, as shown in Figure~\ref{fig:sparsity_matrix}. With the remote part containing the matrix elements that interact with the remote part of the vector, the system matrix is effectively transformed to an irregular matrix and the selection of DIA format would result in excessive zero-padding and potentially run out of memory. In addition to changing the regularity of the matrix, note that from square matrix it is now rectangular. 

We measure the single node performance (i.e. with MPI disabled) of each supported format, over a number of problem sizes, systems and architectures in order to determine the performance benefit from switching to the best available format (see Section~\ref{sec:single-node-perf}). In addition, in Section~\ref{sec::multi-node-perf} we split the local and remote matrix parts and by switching to the available format combinations for each part we evaluate the scaling performance application with respect to the performance of the reference HPCG implementation.

\subsection{Compute node architectures}
All experiments were carried out on the ARCHER2 and Cirrus supercomputers; their compute node architectures are described in Table~\ref{tab:platforms}. Note that Cirrus has both CPU-only as well as CPU+GPU nodes. Also note that for each experiment we explicitly state the compilers used, however the compiler optimization flag \verb|-O3| is always used. For each experiment build, run and data processing scripts for the machines used are available online.\footnote{Available at: https://github.com/morpheus-org/morpheus-benchmarks.git}
\begin{table}[h]
\vskip 3mm
\begin{center}
\begin{small}
\begin{sc}
\begin{tabular}{lccc}
\hline
Platform    &   Cirrus          &   Cirrus          &   ARCHER2     \\
            &   (GPU Node)      &   (CPU Node)      &               \\
\hline
            &   Intel           & Intel             &    AMD     \\
 CPU        & Xeon Gold         &   Xeon            &   EPYC   \\
            & 6248 (x2)         &   E5-2695 (x2)    &   7742 (x2)      \\
\hline
            & NVIDIA            &           &           \\
GPU         & Tesla V100        &   N/A     &   N/A     \\
            & SXM2-16GB (x4)    &           &           \\
\hline
\end{tabular}
\end{sc}
\end{small}
\setlength{\belowcaptionskip}{-8pt}   
\caption{Node configurations for the systems used in the experiments.}
\label{tab:platforms}
\end{center}
\vskip -3mm
\end{table}

\subsection{Overhead Comparison}\label{sec:overheads}
It is vital for an abstraction framework to not introduce overheads that might negate any performance gains. In order to quantify any overheads introduced by \emph{Morpheus} we compare the performance of \gls{spmv} as used by the \emph{Morpheus}-HPCG (with \emph{DynamicMatrix} set to CSR) to the original HPCG implementation, both using the same backends (Serial/OpenMP) on ARCHER2. The experiment is repeated over a set of per-core problem sizes. Note that for the OpenMP backend we are limiting the experiment to a single NUMA region (16 cores on AMD EPYC 7742) to avoid any NUMA effects.

The efficiency of the dynamic dispatch mechanism used by \emph{Morpheus} is implementation specific and overheads might vary across compilers. We therefore repeated the experiments across all available compilers on ARCHER2: GNU 9.3.0, 10.3.0 and 11.2.0; AOCC 2.2.0 and 3.0.0; and CRAY-clang 11.0.4 and 12.0.3.

\begin{figure}[t]
    \centering
    \includegraphics[width=0.9\columnwidth]{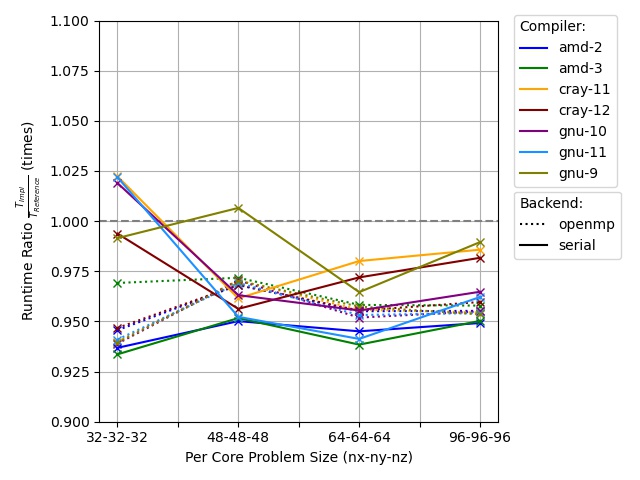}
    \setlength{\belowcaptionskip}{-8pt} 
    \caption{Overheads resulting from adding \emph{Morpheus} to HPCG. We compute the runtime ratio of SpMV as computed by the \emph{Morpheus}-HPCG (\emph{DynamicMatrix} set to CSR) and the original HPCG implementation on ARCHER2. A ratio above $1$ indicates that overheads are introduced, a ratio below $1$ indicates a performance improvement.}
    \label{fig:overheads}
\end{figure}

Figure~\ref{fig:overheads} shows that the \gls{spmv} runtime ratio of the \emph{Morpheus}-HPCG and the the original implementation is concentrated slightly below 1. This indicates that no significant overheads are introduced by \emph{Morpheus} for both the Serial and OpenMP backends. On the contrary, the runtime is slightly improved ($\approx5\%$). The slight improvement in performance can be attributed to the fact that the original implementation uses a variation of CSR as the format of choice, where the values and column indices are accessed by traversing pointers-of-pointers, whilst the \emph{Morpheus}-enabled version uses the traditional CSR.

\begin{figure*}
    \centering
    \captionsetup{justification=centering}
    \begin{subfigure}[h]{0.42\textwidth}
            \captionsetup{justification=centering}
            \includegraphics[width=\columnwidth]{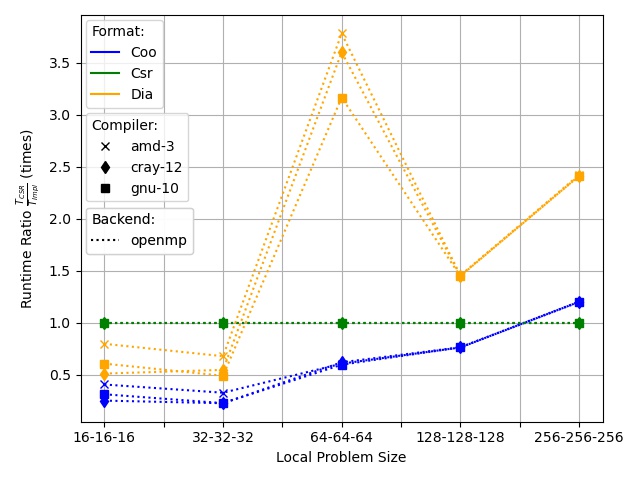}
            \caption{ARCHER2}
            \label{fig:single-node-archer2}
    \end{subfigure}
    ~
    \begin{subfigure}[h]{0.42\textwidth}
            \captionsetup{justification=centering}
            \includegraphics[width=\columnwidth]{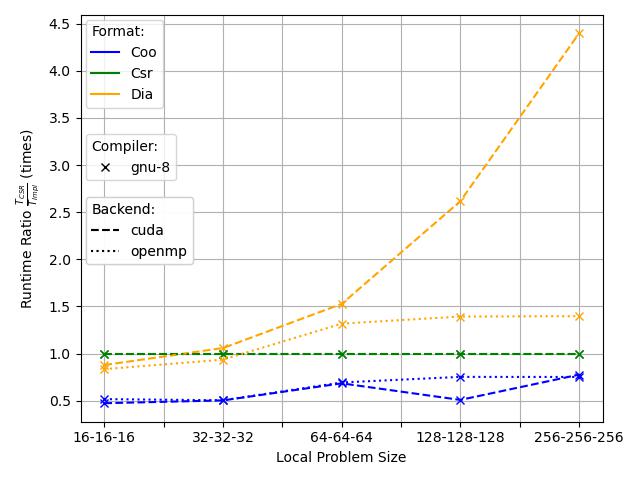}
            \caption{Cirrus}
            \label{fig:single-node-cirrus}
    \end{subfigure}
    \caption{Single node performance of the \emph{Morpheus}-HPCG. For each backend and compiler, the performance is measured as the \gls{spmv} runtime ratio of the \emph{DynamicMatrix} with active state set to CSR w.r.t the \emph{DynamicMatrix} with an active state to each of the supported storage formats (COO, CSR, DIA) over a set of problem sizes on ARCHER2 and Cirrus. A ratio above $1$ indicates a speedup over the performance achieved when using CSR.}
    \label{fig:single-node}
    
    \begin{subfigure}[h]{0.42\textwidth}
            \captionsetup{justification=centering}
            \includegraphics[width=\columnwidth]{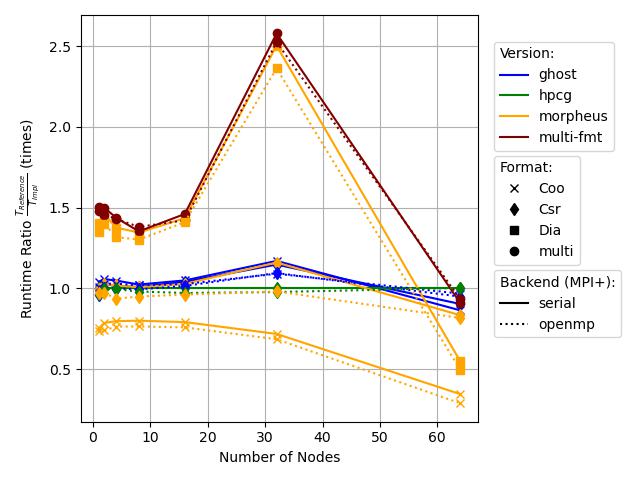}
            \caption{Strong Scaling (ARCHER2)}
            \label{fig:strong-archer2}
    \end{subfigure}
    ~
    \begin{subfigure}[h]{0.42\textwidth}
            \captionsetup{justification=centering}
            \includegraphics[width=\columnwidth]{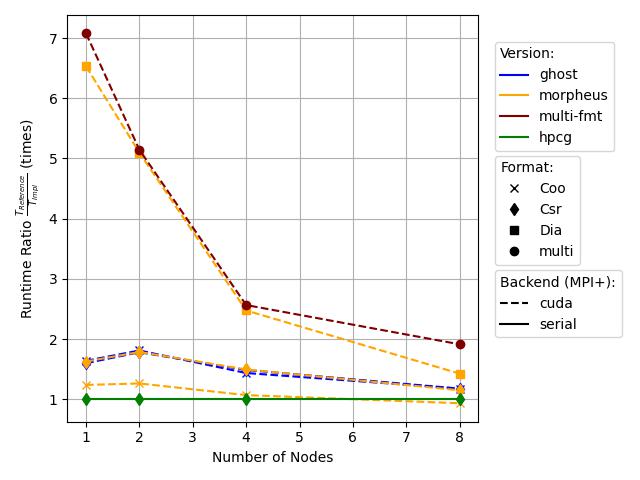}
            \caption{Strong Scaling (Cirrus)}
            \label{fig:strong-cirrus}
    \end{subfigure}

        \begin{subfigure}[h]{0.42\textwidth}
            \captionsetup{justification=centering}
            \includegraphics[width=\columnwidth]{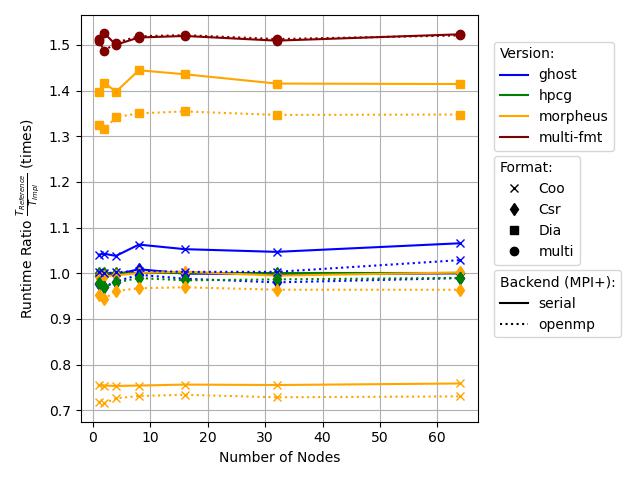}
            \caption{Weak Scaling (ARCHER2)}
            \label{fig:weak-archer2}
    \end{subfigure}
    ~
    \begin{subfigure}[h]{0.42\textwidth}
            \captionsetup{justification=centering}
            \includegraphics[width=\columnwidth]{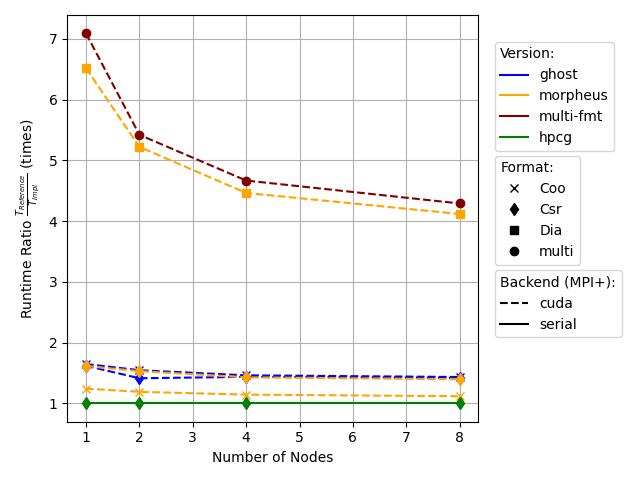}
            \caption{Weak Scaling (Cirrus)}
            \label{fig:weak-cirrus}
    \end{subfigure}
    \caption{Multi-node strong and weak scaling performance of \emph{Morpheus}-HPCG. The \gls{spmv} runtime ratio for each number of nodes(/GPUs) is obtained by evaluating the MPI-only reference HPCG implementation w.r.t to the versions provided by \emph{Morpheus}-enabled HPCG. Available versions are: 1)~Original HPCG 2)~Morpheus (Local part changes format, remote part in CSR) 3)~Ghost (Local part in CSR, remote part changes) 4)~Multi-format (Local and remote part change per process to the optimum format).}
    \label{fig:multi-node-performance}
\end{figure*}

\subsection{Single Node Performance}\label{sec:single-node-perf}
By integrating \emph{Morpheus} into an application users can take advantage of its dynamic switching capabilities and target heterogeneous environments. In this experiment, we show the performance gains that can be achieved  by changing the active type of the \emph{DynamicMatrix}.

In order to quantify the performance gain, we run the \emph{Morpheus}-HPCG with MPI disabled and measure the \gls{spmv} runtime. For each run, we change the active state of the \emph{DynamicMatrix} to one of the supported formats (i.e. COO, CSR, DIA). The runtime ratio is computed by dividing the runtime of the reference state (CSR) with the actual state we are measuring for. The experiment is repeated for a set of problem sizes, available compilers and backends, both on ARCHER2 and Cirrus.

On ARCHER2, one compiler from each vendor was used (GNU~10.3.0, AOCC~3.0.0 and CRAY-clang~12.0.3) and the backend of choice was OpenMP. To avoid NUMA effects, only a single chiplet (16 cores) was used. On Cirrus the experiment was repeated on both CPU and GPU nodes. On the CPU nodes, the compiler of choice was GNU~8.2.0 with the OpenMP backend enabled, and a single CPU (18 cores) was used, again to avoid NUMA effects. On the GPU nodes the device code was compiled using NVCC~11.6 and the host code with GNU~8.2.0, with CUDA backend enabled and the experiment was run on a single GPU.

Figure~\ref{fig:single-node} confirms our hypothesis that the DIA format would perform best both on CPU and GPU, giving improvements of $3.5\times$ on ARCHER2 and up to $4.5\times$ on a Cirrus GPU node compared to the equivalent CSR implementations using the same backend. Since the matrix is highly regular with non-zeros around the diagonals, the DIA format exploits this by offering direct and contiguous access to both the values of the matrix $A$ and the vector $x$, eliminating the need for indirect accesses. However for smaller problem sizes (up to $32\times32\times32$ - see Figures~\ref{fig:single-node-archer2} and \ref{fig:single-node-cirrus}), CSR is either performing better or on par with DIA as the system matrix is small enough such that the costs from indirection are less than the ones from the zero-padding introduced by DIA. In addition, even though COO underperforms compared to CSR, it is worth pointing out that for a problem size of $256\times256\times256$ on ARCHER2, COO now beats CSR performance. This observation further motivates the use of dynamic matrices and shows the importance of having the ability to adapt the data-structure at runtime.

\subsection{Multi-node Performance}\label{sec::multi-node-perf}
Enabling MPI results in an unstructured local system matrix with a sparsity pattern similar to Figure~\ref{fig:sparsity_matrix}. As a result, DIA introduces excessive zero padding and will cause out-of-memory errors for large problems. Instead of conceptually dividing the local system matrix into local and remote parts, we can actually split it into two parts and convert each part to \emph{DynamicMatrix}. By exploiting the runtime switching functionality supported by \emph{Morpheus} we can change the active state of each matrix to potentially different formats based on the sparsity pattern, and also extend this principle across processes.

In this experiment we investigate the multi-node scaling performance of the different versions of \emph{Morpheus}-enabled HPCG. Scaling performance is evaluated both in terms of strong scaling (increasing number of processing units whilst keeping the same global problem) and weak scaling (maintaining same problem size per node whilst increasing number of processing units). The reference version is the original MPI-enabled HPCG and the \emph{Morpheus}-enabled versions are: 1)~\emph{Morpheus}, where the storage format of the local matrix varies for all processes in the same way and the remote matrix is set to CSR 2)~\emph{Ghost}, where the remote storage format varies for all processes and local matrix is set to CSR and 3)~\emph{Multi-Format}, where the formats of both local and remote matrices vary and can be different across processes. To obtain the runtime ratio we divide the \gls{spmv} runtime of the reference MPI-enabled HPCG by the \gls{spmv} runtime of each version. A ratio above $1$ indicates a speed-up over the performance achieved with the original HPCG implementation.

The experiments on ARCHER2 for the Serial and OpenMP backends are each configured with 128 processes and 8 processes each with 16 threads respectively. The global problem size for strong scaling is $512\times512\times256$ and the same problem size is used per node for the weak scaling. For Cirrus we use the GPU nodes to benchmark the CUDA backend, which is configured as one GPU per process. For comparison, we evaluate the MPI-only reference HPCG on the CPU nodes. The global problem size for strong scaling is set to $384\times256\times128$, and we duplicate this per node during the weak scaling experiment. Note that the CPU cores on Cirrus are underpopulated (32 processes out of the available 36) to simplify the computation process of problem size as we vary resources.

In order to select the best format for each process and matrix in the \emph{Multi-Format} version, a naive auto-tuner is used. For each combination of formats, profiling runs are performed and the per-process runtime of the \gls{spmv} kernel is recorded. The auto-tuner then selects the best performing format combination for each process.

Figure~\ref{fig:strong-archer2} shows that on ARCHER2, \emph{Morpheus}-enabled HPCG runs up to $2.5\times$ faster when the active state of the local matrix is set to DIA compared to the MPI-enabled reference HPCG. This remains true until the problem size per node becomes very small; this happens at $64$ nodes when the system matrix becomes almost dense, and now the CSR format outperforms DIA by $2\times$. In addition, changing the active state of the remote matrix slightly improves runtime with up to $1.2\times$ when it is set to COO. However, by adopting the multi-format approach, as the amount of resources varies and different formats perform best, we are now able to maintain the best possible performance since we are no longer restricted to using a single storage format. Similarly, in Figure~\ref{fig:strong-cirrus} the GPU runtime using DIA compared to the MPI-enabled reference HPCG is $6.5\times$ faster on a single GPU and drops to $1.3\times$ on 8 GPUs. This can be attributed to the fact that the problem size per GPU significantly reduces and consequently occupancy on each GPU reduces too. Furthermore, as the number of GPUs increases, the time spent in the \texttt{ExchangeHalo} routine increases and approaches the times spent in the \gls{spmv} kernel. The multi-format version achieves similar performance as the \emph{Morpheus} version and outperforms it by $25\%$ on 8 GPUs as it sets the active type of the local matrix to DIA and the ghost matrix on half of the processes to COO and the other half to CSR. Note that because now the local and remote matrices on each process are separate, it is also possible to overlap communication and computation in order to reduce communication time and improve scaling, however this is out of the scope of this work.

The weak scaling experiments in Figures~\ref{fig:weak-archer2} and \ref{fig:weak-cirrus} show that for the current problem size per node the best performance is obtained when the active state of the local matrix is set to DIA. Changing the active state of the remote matrix from CSR to COO will marginally improve runtime. For ARCHER2, it is worth pointing out that the optimal setup (which improves runtime by $1.5\times$) is obtained when the active states of local and remote matrices on each process are set to DIA and COO respectively. On Cirrus, weak scaling demonstrates similar trends as strong scaling however because the load per GPU here is larger and the communication times are the same as before, the runtime performance improvement saturates at $4\times$. Note that for both architectures and available backends, in the case where the problem size per node is very small, the optimal format ends up being CSR for both the local and remote matrices.
\section{Related Work}
Research efforts into optimizing sparse computations focus on many directions with the most prominent being the creation of novel storage formats, and performance tuning of operations through parameter optimization or automatic format selection.

\textbf{Storage Formats} Many of the new formats that are proposed are derivations or extensions of existing formats aiming to improve performance by addressing known weaknesses. In addition, hybrid approaches that combine multiple formats into a single data structure have been proposed, with the aim to better exploit certain sparsity patterns. Each hybrid format uses mechanisms to determine in which underlying format each portion of the matrix is stored. Examples include \gls{hyb}\cite{monakov_ellpack} format, that combines \gls{ell}\cite{spmv_gpu_review} and \gls{coo}, \gls{hdc}\cite{hdc} and \gls{hec}\cite{hec}. The cocktail\cite{clSpMV} format takes the idea a step further and partitions the matrix into many sub-matrices by collecting matrix features and enforcing partition policies with each partition assigned one of the nine available formats.

\textbf{Auto-tuners} Auto-tuners such as OSKI\cite{oski}, FFTW\cite{fftw} and ATLAS\cite{atlas} are few of the successful approaches in domain-specific performance-critical libraries. In the field of sparse linear algebra, the focus of auto-tuners is mainly on tuning performance through selecting the storage format, the kernel implementation or parameters. For example, SMAT\cite{smat}, SMATER\cite{smater}, clSpMV\cite{clSpMV} and Zhao et al.\cite{cnn_spmv} enable runtime selection of the best format and \gls{spmv} kernel implementation across architectures for a given matrix. In a similar manner, Xie et al.\cite{ia-spgemm} implement an auto-tunner that automatically determines the best format and \gls{spgemm} algorithm for arbitrary sparse matrices.

\textbf{Libraries} PETSc\cite{petsc}, Eigen\cite{eigen}, AMGX\cite{amgx} and CUSP\cite{cusp} are only few of the examples of software libraries developed for solving linear systems. Although these solutions constitute the current state-of-the-art, they are either specific to a single target hardware or support only a single format internally and it is often a significant undertaking to provide full support for a new format. Given the heterogeneous nature of modern \gls{hpc} systems, providing support for multiple target architectures and provisions for supporting future architectures is a requirement. 

\section{Conclusions and Further Work}
As there is no single sparse matrix storage format that will perform optimally across different operations, sparsity patterns and target architectures, the ability to dynamically change the underlying data-structure to better match the computational pattern and hardware characteristics enables a range of optimization opportunities. Extending this concept into a distributed environment allows us to create a dynamic distributed matrix capable of changing its storage format on each process and achieve the best possible performance irrespective of the number or type of resources that are available.

Providing end-users with a mechanism to 
optimize their applications without any code changes (beyond the initial porting effort) and allowing them to target new architectures effectively increases the life-time of their software as support for more formats and target architectures is added.

As a next step, we will add further storage formats to \emph{Morpheus} and study the performance characteristics over a wider set of matrices with different sparsity patterns. In addition, updating the auto-tuner to automatically select the optimum format online and not from profiling runs remains an avenue for further research.
\section*{Acknowledgment}
This research is part of the EPSRC project ASiMoV (EP/S005072/1). We used the ARCHER2 UK National Supercomputing Service (https://www.archer2.ac.uk) and the Cirrus UK National Tier-2 HPC Service at EPCC, funded by the University of Edinburgh and EPSRC (EP/P020267/1).

\bibliographystyle{bibliography/IEEEtran}  
\bibliography{bibliography/bibs.bib}

\end{document}